\documentclass[epj,referee]{svjour}
\usepackage{graphicx}
\usepackage{xspace}

\usepackage{amsmath,amssymb}
\usepackage[T1]{fontenc}
\usepackage[latin9]{inputenc}
\usepackage{amstext}
\usepackage{esint}

\begin{document}

\title{Cold atom-molecule photoassociation: long-range interactions beyond the $1/R^{n}$ expansion}

\author{M. Lepers \and O. Dulieu}

\institute{Laboratoire Aim\'e Cotton, CNRS, B\^at. 505, Univ Paris-Sud, 91405 Orsay Cedex, France}
\date{\today}

\abstract{
We investigate theoretically the combination of first-order quadrupole-quadrupole and second-order dipole-dipole effects on the long-range electrostatic interactions between a ground-state homonuclear alkali-metal dimer and an excited alkali-metal atom. As the electrostatic energy is comparable to the dimer rotational structure, we develop a general description of the long-range interactions in the framework of the second-order degenerate perturbation theory, which allows for couplings between the dimer rotational levels. The resulting adiabatic potential energy curves exhibit avoided crossings, and cannot be expanded on the usual $1/R^{n}$ series. We study in details the breakdown of this approximation in the particular case Cs$_{2}$+Cs($6^2P$). Our results are found promising to achieve photoassociation of ultracold trimers.
}

\maketitle

\section{Introduction}
\label{sec:intro}
Researches on cold ($T \lesssim 1$~K) and ultracold ($T \lesssim 1$~mK) gases have attracted a considerable interest over the last years, as at such temperatures, the properties of the gases are predominantly governed by purely quantum effects, like the tunneling, quantum resonances, and quantum Bose and Fermi statistics. One of the most recent developments concerns cold and ultracold molecules -- the central theme of the present topical issue \cite{dulieu2009,carr2009}. The exquisite control of both external and internal degrees of freedom of the molecules offers amazing opportunities for fundamental and applied research, like high-resolution molecular spectroscopy \cite{claussen2003,kemmann2004,vanhaecke2004,wang2004}, resonant dynamics \cite{boesten1997,lang2008,chin2010}, phase-driven chemistry \cite{heinzen2000,donley2002}, quantum degeneracy \cite{greiner2003,jochim2003,zwierlein2003}, quantum phase transitions \cite{bartenstein2004,bourdel2004,chin2004a}, universal quantum states \cite{braaten2006,kraemer2006,ferlaino2008}, metrology of fundamental constants \cite{demille2000,hudson2002,schiller2005,chin2006}, quantum information \cite{demille2002,yelin2006,charron2007}, cold and ultracold collisions and chemistry \cite{elioff2003,zahzam2006,staanum2006,hudson2008,levinsen2009,knoop2010,ospelkaus2010a,kirste2010}.

The central feature of cold and ultracold gases is that the relative kinetic energy of the particles is so tiny that their interactions is dominated by long-range forces induced by the spatial extension of their charge distribution. The related potential energy is usually described by the well-known multipolar expansion $\sum_{n}C_{n}/R^{n}$ in inverse powers of the (large) distance $R$ between the interacting particles \cite{fontana1961a,fontana1961b,fontana1962}. With the development of ultracold physics, a vast amount of accurate calculations of the $C_n$ coefficients is available nowadays for several atomic species like alkali-metal \cite{bussery1987,marinescu1995,marinescu1999,derevianko2001,porsev2003,mitroy2003}, alkaline-earth \cite{porsev2002}, rare gas atoms \cite{derevianko2000}, or open-shell atoms \cite{rerat1997,gronenboom2007}.

In the present study, as the third of a series of papers \cite{lepers2010,lepers2011} (hereafter referred to as papers I and II, respectively), we investigate the long-range interaction of an excited alkali-metal atom and a ground state alkali-metal diatomic molecule. Indeed, samples of ultracold ground state alkali-metal diatomics are available in several experiments, surrounded in most cases by ultracold atoms \cite{haimberger2004,sage2005,pechkis2007,viteau2008,deiglmayr2008,danzl2008,ni2008,lang2008a}. Tuning the frequency $\nu$ of a laser to the red of a resonant atomic transition, it should be possible to induce the photoassociation (PA) of an atom $\textrm{A}$ and a molecule $\textrm{BC}$ according to $\textrm{A}+\textrm{BC}+h\nu \rightarrow (\textrm{ABC})^*$ just like it is routinely achieved for atoms \cite{jones2006}. The excited complex $\textrm{ABC}^*$ could spontaneously decay afterwards in a stable ultracold trimer $\textrm{ABC}$, providing then an example of a photo-assisted ultracold chemical reaction. While the latter process most probably relies on complex chemical forces, the former PA step is obviously controlled by the long-range interaction between the excited atom $\textrm{A}^*$ and the $\textrm{BC}$ molecule. In contrast with the atom-atom case, the long-range interactions between atoms and molecules has been less often addressed quantitatively in the literature. In most studies, they have been evaluated at fixed geometries in order to match or to fit at large distances potential energy surfaces obtained by quantum chemistry computations \cite{rerat2003,merawa2003,bussery-honvault2008,bussery-honvault2009,berteloite2010}. Note however that in a recent paper \cite{kotochigova2010a}, Kotochigova reported a study on the van der Waals interaction between a ground state alkali metal atom and a ground state alkali-metal diatomic in a given rovibrational level, \text{i.e.} beyond the fixed geometry framework. As an illustration, we focus here on the PA reaction between an ultracold ground state Cs$_2$ molecule in a vibrational level $v_d$ and a rotational level $N$ and a cesium atom
\begin{equation}
\textrm{Cs}_{2}(X^{1}\Sigma_{g}^{+},v_{d},N)+\textrm{Cs}(6^{2}S)+h\nu \to\textrm{Cs}_{3}^{*}
\label{eq:reaction-Cs3}
\end{equation}
which could be explored in the experimental group in Orsay. In paper I we have determined the leading term of the long-range interaction, namely the quadrupole-quadrupole term between an excited Cs($6^{2}P$) atom and a ground state Cs$_{2}(X^{1}\Sigma_{g}^{+},v_{d},N)$ molecule, with $N >0$. We extended this study in paper II to the calculation of van der Waals coefficients with second-order perturbation theory. Due to the competition between the rotational energy of the molecule and the electrostatic interactions, we demonstrated that the validity of the multipolar expansion at short distances is not limited by the overlap between electronic clouds of the particles (occurring around 45~a.u. in the present case), but by the degeneracy between various long-range states of the complex occurring in the 80-100~a.u. distance range (1~a.u.=$a_0$=0.0529177~nm). The aim of this article is therefore to investigate the distance range for which the multipolar expansion is still applicable to compute long-range interactions, but resulting into potential curves determined by a second-order degenerate perturbation theory which cannot be written anymore as the usual $\sum_{n}C_{n}/R^{n}$ expression. The paper is organized as follows. In Section \ref{sec:basics}, we recall the basics of the perturbative calculations of long-range interactions, as described in papers I and II, and reformulate it up to the level of the second-order degenerate perturbation theory. In Section \ref{sec:theory}, we derive the expressions of the first-order and second-order interaction energy. The results for the interaction of a ground state Cs$_2$ molecule and an excited Cs atom are presented in Section \ref{sec:results} as the initial state of the photoassociation of a cold atom-molecule pair, before addressing some prospects in Section \ref{sec:conclusion}.

\section{Energy scales and perturbation theory}
\label{sec:basics}
\begin{figure}
\begin{centering}
\includegraphics[scale=0.4]{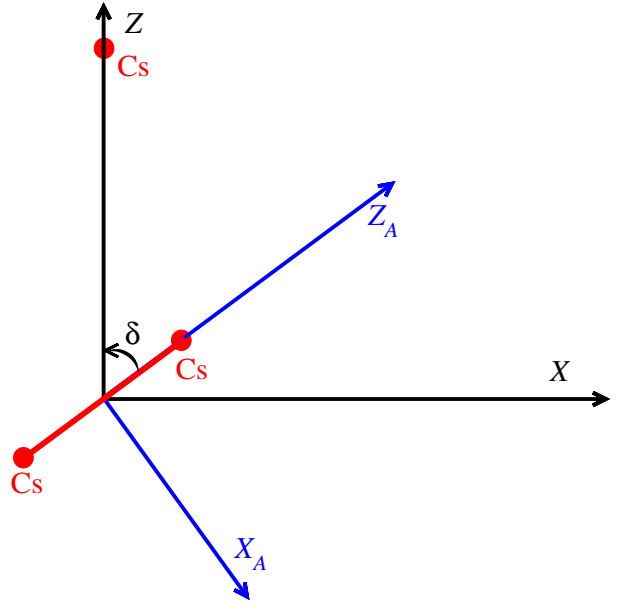}
\par\end{centering}
\caption{\label{fig:CS}The two coordinate systems, $X_{A}Y_{A}Z_{A}$ (D-CS) and $XYZ$ (T-CS) defined for the dimer and for the trimer, respectively. The $Z_{A}$ axis is along the dimer axis, while $Z$ is oriented from the center of mass of the dimer towards the atom $B$. The $Y$ and $Y_{A}$ axes coincide and point into the plane of the figure. The subsystem $A$ in this figure is the Cs$_{2}$ molecule, the subsystem $B$ is the Cs atom. The T-CS is related to the laboratory coordinate system ($\tilde{x}\tilde{y}\tilde{z}$) by the usual Euler angles ($\alpha,\beta,\gamma$), not represented here.}
\end{figure}
In our two previous articles I and II, the long-range interactions between a homonuclear alkali-metal dimer and an excited alkali-metal atom is investigated, choosing the example of Cs$_{2}+$Cs$^*$. The dimer is in its ground electronic state $\left|X^{1}\Sigma_{g}^{+}\right\rangle $, in an arbitrary vibrational state $\left|v_{d}\right\rangle $, and rotational state characterized by the quantum number $N$ with projection $m$ on the $Z$ axis of the trimer coordinate system (T-CS) (Fig. \ref{fig:CS}). The alkali-metal atom is in its first excited state $^{2}P$ characterized with general indexes labeling  the valence electron: $n$ for the principal quantum number, $\ell$ for the orbital one and its projection $\lambda$ on the $Z$ axis. The fine structure of the atomic level will be included in a further study.

We started from the general form of the electrostatic energy between two interacting charge distributions $A$ (the dimer) and $B$ (the atom), separated by the distance $R$, written as the well-known multipolar expansion:
\begin{eqnarray}
\hat{V}_{el}(R) & = & \sum_{L_{A},L_{B}=0}^{+\infty}\sum_{M=-L_{<}}^{L_{<}}\frac{1}{R^{1+L_{A}+L_{B}}}\nonumber \\
 & \times & f_{L_{A}L_{B}M}\hat{Q}_{L_{A}}^{M}(\hat{r}_{A})\hat{Q}_{L_{B}}^{-M}(\hat{r}_{B}),
\label{eq:MultipoleExp}
\end{eqnarray}
where $L_{<}$ is the minimum of $L_{A}$ and $L_{B}$. Each multipole of order $L_{X}$ (with $X=A,B$) is associated to the tensor operator $\hat{Q}_{L_{X}}^{M}(\hat{r}_{X})$, which can be expressed in a coordinate system whose origin is the center of mass of $X$:
\begin{equation}
\hat{Q}_{L_{X}}^{M}(\hat{r}_{X})=\sqrt{\frac{4\pi}{2L_{X}+1}}\sum_{i\in X}q_{i}\hat{r}_{i}^{L_{X}}Y_{L_{X}}^{M}\left(\hat{\theta}_{i},\hat{\phi}_{i}\right),
\label{eq:OperMultip}
\end{equation}
where $q_{i}$ is the value of each charge $i$ composing $X$, and $Y_{L_{X}}^{M}$ are the usual spherical harmonics. In Eq. (\ref{eq:MultipoleExp}), the assumption has been made that the quantization axis is $Z$, oriented from the center of mass of $A$ to the center of mass of $B$, hence the factor $f_{L_{A}L_{B}M}$ reads
\begin{eqnarray}
f_{L_{A}L_{B}M} & = & \frac{\left(-1\right)^{L_{B}}\left(L_{A}+L_{B}\right)!}{\sqrt{\left(L_{A}+M\right)!\left(L_{A}-M\right)!}} \nonumber \\
 & \times & \frac{1}{\sqrt{\left(L_{B}+M\right)!\left(L_{B}-M\right)!}}.
\end{eqnarray}
In Papers I and II, we defined the zeroth-order energy $\mathcal{E}_{p}^{0}$ of a given $(2N+1)(2\ell+1)$ times degenerate state labelled $p$ as
\begin{equation}
\mathcal{E}_{p}^{0}=E_{Xv_{d}N}+E_{n\ell}
\end{equation}
where $E_{n\ell}$ is the energy of the free atom, $E_{Xv_{d}N}=E_{Xv_{d}}+B_{Xv_{d}}N\left(N+1\right)$ the free energy of the dimer in the rovibrational level ($v_d,N$) considered for simplicity sake as a rigid rotor with a rotational constant $B_{Xv_{d}}$ ($B_{Xv_d=0}=B_0=1.173\times10^{-2}$~cm$^{-1}$ \cite{amiot2002a}). In the following, the energy origin will be set at $E_{X,v_{d}=0}+E_{6P}=0$ without loss of generality of our treatment. As the two reactants get closer from each other, the degeneracy over the different sublevels is progressively lifted, and the electrostatic energy must be calculated using the perturbation theory for degenerate levels.

The first-order correction is the quadrupole-quadrupole energy of interaction $\hat{V}_{qq}$ scaling as $R^{-5}$ obtained by setting $L_{A}=L_{B}=2$ in Eq. (\ref{eq:MultipoleExp})
\begin{equation}
\hat{V}_{qq}=\frac{24}{R^{5}}\sum_{M}\frac{\hat{Q}_{2}^{M}(\hat{r}_{A})\hat{Q}_{2}^{-M}(\hat{r}_{B})}{\left(1+M\right)!\left(1-M\right)!}\:.
\label{eq:Vqq}
\end{equation}
The corresponding potential energy curves $B_{0}N(N+1)+C_{5}/R^{5}$ as functions of $R$ have been reported in paper I. In the case of Cs$_{2}+$Cs$^*$, the  curves start to cross each other at distances $R_{m}\approx100$ a.u.. This distance is higher than the Leroy radius \cite{leroy1974} estimated at $R_{LR}\approx45$ a.u. where electronic clouds start to overlap. Therefore we define the crossing region by $R_{LR}\lesssim R\lesssim R_{m}$  where the perturbation approach described previously is not applicable. This is a general feature due to the competition of the rotational energy of the dimer with the electrosatic energy. Note that for Li$_{2}+$Li, we estimate this crossing in the $26\lesssim R\lesssim43$~a.u. range.

In paper II, we studied the second-order contribution to the dipole-dipole interaction, i.e. $L_{A}=L_{B}=1$, scaling as $R^{-6}$. We calculated
the $C_{6}$ coefficient corresponding to each eigenvector $\left|\varphi_{p}^{0}\right\rangle $ of $\hat{V}_{qq}$ in the subspace of degeneracy associated to $\mathcal{E}_{p}^{0}$. Noting that
\begin{equation}
\left|\varphi_{p}^{0}\right\rangle =\sum_{m\lambda}c_{m\lambda}\left|m\lambda\right\rangle ,
\end{equation}
we write the $C_{6}$ coefficient in the general form
\begin{equation}
C_{6}=R^{6}\sum_{m_{1}\lambda_{1}}\sum_{m_{2}\lambda_{2}}c_{m_{1}\lambda_{1}}c_{m_{2}\lambda_{2}}\left\langle m_{1}\lambda_{1}\right|\hat{V}_{dd}^{(2)}\left|m_{2}\lambda_{2}\right\rangle ,
\end{equation}
where $\hat{V}_{dd}^{(2)}$ is the second-order dipole-dipole operator
\begin{eqnarray}
\hat{V}_{dd}^{(2)} & = &
-4\sum_{a,b}\sum_{M,M'}\frac{1}{\left(E_{a}-E_{Xv_{d}N}\right)+\left(E_{b}-E_{n\ell}\right)}\nonumber \\
 & \times & \frac{\hat{Q}_{1}^{M}\left|\Phi_{a}\right\rangle
\left\langle \Phi_{a}\right|\hat{Q}_{1}^{-M'}
\hat{Q}_{1}^{-M}\left|\Phi_{b}\right\rangle
\left\langle\Phi_{b}\right|\hat{Q}_{1}^{M'}}
{\left(1+M\right)!\left(1-M\right)!\left(1+M'\right)!\left(1-M'\right)!}\,,
\label{eq:Vdd2}
\end{eqnarray}
where $a\equiv X'v'_{d}N'm'$ and $b\equiv n'\ell'\lambda'$ are the quantum numbers associated to the intermediate states accessible by
dipolar transition. In paper II, the calculations for Cs$_{2}+$Cs$^*$ revealed that, in the crossing region, the second-order dipolar energy
becomes comparable to the first-order quadrupolar and the rotational energies.

In order to describe the electrostatic interactions in the crossing region, we need to formulate the perturbation approach in a different way. As the rotational and electrostatic energies are comparable, we consider the perturbation hamiltonian $\hat{W}$
\begin{equation}
\hat{W}=B_{Xv_{d}}\hat{\vec{N}}^{2}+\hat{V}_{qq}+\hat{V}_{dd}^{(2)}.
\label{eq:HamiltPertub}
\end{equation}
so that the zeroth-order energy of a state $p$ is reduced to $E_{p}^{0}=E_{Xv_{d}}+E_{n\ell}$. Each value of $E_{p}^{0}$ is associated to a subspace of quasi-degeneracy whose basis vectors can be labeled by $N$, $m$ and $\lambda$ quantum numbers. In the present formulation, the electrostatic
interactions now couple different values of $N$, so that each subspace of degeneracy has in principle an infinite dimension.

It may seem surprising that, in Eq. (\ref{eq:HamiltPertub}), $V_{dd}^{(2)}$ is actually manipulated like a first-order contribution, whereas it is a second-order term. To justify Eq. (\ref{eq:HamiltPertub}), we have calculated the second-order correction to energies and state vectors for degenerate unperturbed levels, and found that the corrections to energy are given by the eigenvalues of the operator $\hat{V}_{dd}^{(2)}$. In principle, in order to calculate the total second-order contribution of the electrostatic energy, dipole-quadrupole, quadrupole-dipole, quadrupole-quadrupole, ... terms must be taken into account. However one should keep in mind that those interactions scale as $R^{-8}$, $R^{-8}$ and $R^{-10}$ respectively. If we refer to the values of $C_{8}$ and $C_{10}$ coefficients (see \textit{e.g.} Ref.\cite{marinescu1995}), we can expect those interactions to be significantly smaller (at least one order of magnitude), in the range of distance that we consider here.

\section{Calculation of the electrostatic energy}
\label{sec:theory}
This section is devoted to the calculations of the matrix elements
of $\hat{V}_{qq}$ and $\hat{V}_{dd}^{(2)}$ in the basis $\left\{ \left|Nm\lambda\right\rangle \right\} $.
The principle of the calculations has been presented in Papers I and II. So we will only recall the main steps, emphasizing on the most important
difference: the couplings between distinct values of the rotational quantum number $N$ are taken into account.
\subsection{First-order quadrupole-quadrupole interaction}
\label{ssec:theory_1st}
As the atomic part of the matrix elements of $\hat{V}_{qq}$ are unchanged in comparison to Eq. (13) of Paper I, we only focus here on the dimer
part. The wave function $\Psi_{Nm}(\delta)$ associated to the rotational state $\left|Nm\right\rangle $ of the dimer is written in
the T-CS as
\begin{equation}
\Psi_{Nm}(\delta)=\sqrt{\frac{2N+1}{2}}d_{m0}^{N}(\delta),
\end{equation}
where $d_{m0}^{N}(\delta)$ is the reduced Wigner matrix element. The prefactor $\sqrt{(2N+1)/2}$ ensures normalization to unity with respect to $\delta$. Therefore, we can write the matrix element of the quadrupole moment of the dimer as
\begin{eqnarray}
& & \left\langle N_{1}m_{1}\left| \hat{Q}_{2}^{M} \right|N_{2}m_{2}\right\rangle \nonumber\\
&=& \frac{\sqrt{(2N_1+1)(2N_2+1)}}{2}q_{2}^{0} \nonumber\\
& \times & \int_{0}^{\pi}d\delta\sin\delta d_{m_{1}0}^{N_1}(\delta) d_{M0}^{2}(\delta) d_{m_{2}0}^{N_2}(\delta) \nonumber \\
&=& \sqrt{\frac{2N_2+1}{2N_1+1}}
C_{N_{2}020}^{N_{1}0}C_{N_{2}m_{2}2M}^{N_{1}m_{1}}q_{2}^{0} \,,
\label{eq:DimQuadrMom}
\end{eqnarray}
where $q_{2}^{0}$ is the (only non-zero) tensor component of the quadrupole moment of the dimer in its coordinate system (the D-CS). In Eq. (\ref{eq:DimQuadrMom}), we have written the quadrupole moment operator (see Eq.(\ref{eq:OperMultip}) with $L=2$) using the property \footnote{In our case, $\theta=\delta$, whereas $\phi$ is not defined so that we take $\phi=0$.} $e^{-iM\phi}d_{M0}^{L}(\theta)=\sqrt{4\pi/(2L+1)}Y_{L}^{M*}(\theta,\phi)$. Then we have calculated the integral of three Wigner functions \cite{varshalovich1988}, and obtained the Clebsch-Gordan coefficients $C_{a\alpha b\beta}^{c\gamma}$. So the matrix element of $\hat{V}_{qq}$ reads
\begin{eqnarray}
& &\left\langle N_{1}m_{1}\lambda_{1}\left|\hat{V}_{qq}\right|N_{2}m_{2}\lambda_{2}\right\rangle  \nonumber \\
 & = & -24C_{N_{2}020}^{N_{1}0}C_{\ell020}^{\ell0}\frac{q_{2}^{0} \left\langle r_{n\ell}^{2}\right\rangle }{R^{5}}\nonumber \\
& \times & \sum_{M=-2}^{2} \frac{C_{N_{2}m_{2}2M}^{N_{1}m_{1}}C_{\ell\lambda_{2}2-M}^{\ell\lambda_{1}}}{\left(2+M\right)!\left(2-M\right)!}\,.
\label{eq:Vqq-MatElem}
\end{eqnarray}
We recall from Paper I, that for Cs$_{2}$, we estimated $q_{2}^{0}$ at 18.58 a.u., and we calculated $\left\langle r_{6P}^{2}\right\rangle =62.65$~a.u. for cesium using an Hatree-Fock method. Equation (\ref{eq:Vqq-MatElem}) impose strong selection rules: (i) the projection of the total orbital quantum number $m_{J}=m+\lambda$ is conserved; (ii) $N_{1}=N_{2},N_{2}\pm2$, which means that the parity of $N$ is conserved.
\subsection{The second-order dipolar interaction}
\label{ssec:theory_2nd}
Once again, only the contribution from the dimer is modified with respect to the developments of Paper II. In order to calculate $\left\langle N_{1}m_{1}\lambda_{1}\left|\hat{V}_{dd}^{(2)}\right|N_{2}m_{2}\lambda_{2}\right\rangle $,
we concentrate on the calculation of the quantity
\begin{eqnarray}
x(a) & = & \left\langle X\Lambda v_{d}N_{1}m_{1}\lambda_{1}\left|\hat{Q}_{1}^{M}\right|X'\Lambda'v'_{d}N'm'\lambda'\right\rangle \nonumber \\
 & \times & \left\langle X'\Lambda'v'_{d}N'm'\lambda'\left|\hat{Q}_{1}^{-M'}\right|X\Lambda v_{d}N_{2}m_{2}\lambda_{2}\right\rangle ,
\label{eq:Vdd2-x}
\end{eqnarray}
where $a$ denotes in a short way the intermediate state $\left|X'\Lambda'v'_{d}N'm'\lambda'\right\rangle $. For the sake of clarity, we explicitly introduce $\Lambda$ and $\Lambda'$,the projection of the dimer orbital momentum on the $Z_{A}$ axis ($\Lambda=0$ in our particular case). The quantity $x(a)$ is part of the dimer polarizability $\alpha_{MM'}^{m_{1}m_{2}}$ (see Eq. (12) of Paper II) according to the relation
\begin{equation}
\alpha_{MM'}^{m_{1}m_{2}}(z)=2(-1)^{M}\sum_{a}\frac{\left(E_{a}-E_{X\Lambda v_{d}N}\right)x(a)}{\left(E_{a}-E_{X\Lambda v_{d}N}\right)^{2}-z^{2}},
\end{equation}
where $z$ is a real or imaginary frequency. The dipolar matrix element is
\begin{eqnarray}
 &  & \left\langle X\Lambda v_{d}Nm\left|\hat{Q}_{1}^{M}\right|X'\Lambda'v_{d}'N'm'\right\rangle \nonumber \\
 & = & \sum_{\mu=-1}^{1}\left\langle Nm\left|d_{M\mu}^{1}\right|N'm'\right\rangle \left\langle X\Lambda v_{d}\left|\hat{q}_{1}^{\mu}\right|X'\Lambda'v_{d}'\right\rangle ,
\end{eqnarray}
where $\mu$ represents the different possible projections on the $Z_{A}$ axis, and $\hat{q}_{1}^{\mu}$ the corresponding tensor components of the dimer dipole moment. $\mu=0$ is associated to $\Sigma\to\Sigma$ transitions and $\mu=\pm1$ to $\Sigma\to\Pi$ transitions. In a similar way as Eq. (\ref{eq:DimQuadrMom}), and by use of the property \cite{varshalovich1988}
\begin{equation}
C_{a\alpha b\beta}^{c\gamma}=\left(-1\right)^{a-\alpha}\sqrt{\frac{2c+1}{2b+1}}C_{c\gamma a-\alpha}^{b\beta},
\end{equation}
we write the rotational part
\begin{eqnarray}
 & & \left\langle Nm\left|d_{M\mu}^{1}\right|N'm'\right\rangle \nonumber\\
 & = & (-1)^{M+\mu}\sqrt{\frac{2N+1}{2N'+1}}C_{Nm1-M}^{N'm'}C_{N01-\mu}^{N'-\mu}.
\end{eqnarray}
Finally, Eq. (\ref{eq:Vdd2-x}) becomes
\begin{eqnarray}
x(a) & = & (-1)^{M}\frac{\sqrt{\left(2N_{1}+1\right)\left(2N_{2}+1\right)}}{2N'+1} \nonumber\\
 & \times & C_{N_{1}m_{1}1-M}^{N'm'}C_{N_{2}m_{2}1-M'}^{N'm'} \sum_{\mu}C_{N_{1}01-\mu}^{N'-\mu} \nonumber \\
 & \times & C_{N_{2}01-\mu}^{N'-\mu}\left|\left\langle X\Lambda v_{d}\left|\hat{q}_{1}^{\mu}\right|X'\Lambda'v_{d}'\right\rangle \right|^{2},
\end{eqnarray}
and
\begin{eqnarray}
\alpha_{MM'}^{m_{1}m_{2}}(z) & = & \sum_{N'm'}\frac{ \sqrt{\left(2N_{1}+1\right)\left(2N_{2}+1\right)}}{2N'+1} \nonumber \\
 &\times & C_{N_{1}m_{1}1-M}^{N'm'}C_{N_{2}m_{2}1-M'}^{N'm'} \nonumber \\
 & \times & \left[C_{N_{1}010}^{N'0}C_{N_{2}010}^{N'0}\alpha_{\parallel}(z)\right.\nonumber \\
 & & \left.+2C_{N_{1}011}^{N'1}C_{N_{2}011}^{N'1}\alpha_{\bot}(z)\right],
\end{eqnarray}
with $\alpha_{\parallel}$ and $\alpha_{\bot}$ respectively the parallel and perpendicular dipole polarizabilities of the dimer in the $X$ electronic state and $v_{d}$ vibrational level (see Paper II for a description of the polarizability calculation for Cs$_{2}$ and Cs). Therefore, the matrix element of $\hat{V}_{dd}^{(2)}$ looks pretty much like Eq. (32) of Paper II:
\begin{eqnarray}
 &  & \left\langle N_{1}m_{1}\lambda_{1}\left|\hat{V}_{dd}^{(2)}\right|N_{2}m_{2}\lambda_{2}\right\rangle \nonumber \\
 & = & -\sum_{MM'}\sum_{N'm'_{}}\sum_{\ell'\lambda'}\frac{3}{\left(1+M\right)!\left(1-M\right)!\left(1+M'\right)!\left(1-M'\right)!}\nonumber \\
 & \times & \frac{\sqrt{\left(2N_{1}+1\right)\left(2N_{2}+1\right)}}{2N'+1}C_{Nm_{1}1-M}^{N'm'}C_{Nm_{2}1-M'}^{N'm'} \nonumber\\
 & \times & \frac{2\ell+1}{2\ell'+1}C_{\ell\lambda_{1}1M}^{\ell'\lambda'}C_{\ell\lambda_{2}1M'}^{\ell'\lambda'}\nonumber \\
 & \times & \sum_{n'}\left[\frac{2}{\pi}\int_{0}^{+\infty}d\omega
\left(C_{N_{1}010}^{N'0}C_{N_{2}010}^{N'0}\alpha_{\parallel}(i\omega)\right.\right. \nonumber \\
 & & \left. +2C_{N_{1}011}^{N'1}C_{N_{2}011}^{N'1}\alpha_{\bot}(i\omega)\right)\alpha_{n\ell n'\ell'}(i\omega)\nonumber \\
 &  & +4\Theta(-\Delta E_{n'\ell',n\ell})
\left(C_{N_{1}010}^{N'0}C_{N_{2}010}^{N'0}\alpha_{\parallel}(\Delta E_{n'\ell',n\ell}) \right. \nonumber \\
 & & \left.\left. +2C_{N_{1}011}^{N'1}C_{N_{2}011}^{N'1}\alpha_{\bot}(\Delta E_{n'\ell',n\ell})\right)
\left(\mu_{n'\ell',n\ell}\right)^{2}\right]\nonumber \\
 & - & \frac{2}{\pi}\sum_{M}\sum_{N'm'}\frac{1}{\left[\left(1+M\right)!\left(1-M\right)!\right]^{2}} \nonumber\\
 & \times & \frac{\sqrt{\left(2N_{1}+1\right)\left(2N_{2}+1\right)}}{2N'+1}C_{Nm_{1}1M}^{N'm'}C_{Nm_{2}1M}^{N'm'}\delta_{m_{1}m_{2}}\delta_{\lambda_{1}\lambda_{2}}\nonumber \\
 & \times & \int_{0}^{+\infty}d\omega\left(C_{N_{1}010}^{N'0}C_{N_{2}010}^{N'0}\alpha_{\parallel}(i\omega)\right. \nonumber\\
 & & \left. +2C_{N_{1}011}^{N'1}C_{N_{2}011}^{N'1}\alpha_{\bot}(i\omega)\right)\alpha_{c}(i\omega).
\label{eq:Vdd2-MatElem}
\end{eqnarray}
In this equation $\alpha_{n\ell n'\ell'}$ is the state-to-state polarizability corresponding to the $n\ell\to n'\ell'$ atomic transition (see Eq. (22) of Paper II), $\alpha_{c}$ is the core contribution to the atomic polarizability, $\mu_{n'\ell',n\ell}=r_{n'\ell',n\ell}C_{\ell010}^{\ell'0}/\sqrt{3}$
is the atomic transition dipole moment, $\Delta E_{n'\ell',n\ell}=E_{n'\ell'}-E_{n\ell}$ is the atomic excitation energy and $\Theta(x)$ is Heaviside's function, here equal to 1 if $\Delta E_{n'\ell',n\ell}<0$.

Equation (\ref{eq:Vdd2-MatElem}) indicates to us that the selection rules associated to $\hat{V}_{dd}^{(2)}$ are the same as $\hat{V}_{qq}$. Therefore the subspace of degeneracy associated to a given zeroth-order energy $E_{p}^{0}$ can be divided into independent subspaces defined: (i) by the projection of the total orbital momentum $m_{J}=m+\lambda$, and (ii) by the parity of the rotational quantum number $N$.
\section{Results and discussions}
\label{sec:results}
The first-order correction to energy due to the operator $\hat{W}$ is obtained after a diagonalization in each subspace of quasi-degeneracy associated to $E_{p}^{0}$ for each value of $R$. The eigenvalues $E_{p}^{1}(R)$ define potential energy curves with avoided crossings, that will thus be called \emph{adiabatic}. The corresponding eigenvectors will be labeled $\left|(p)\left|m_{J}\right|,N\right\rangle$, where $\left|m_{J}\right|=\left|m+\lambda\right|$, $N$ is the dimer rotational quantum number to which the state tends as $R\to\infty$, and $p$ a number starting from 1 and increasing with energy in a given symmetry (that is to say a given $\left|m_{J}\right|$ and parity of $N$).
By contrast, the potential energy curves and eigenvectors obtained in Papers I and II are called \emph{diabatic}, as they cross inside in a given symmetry. The eigenvectors are labelled $\left|(p)\left|m_{J}\right|,N^{d}\right\rangle $, the $d$ superscript standing for diabatic. Unlike the adiabatic states, the diabatic ones are totally included in the manifold defined by a given value of the dimer rotational quantum number $N$.
\begin{figure}
\begin{centering}
\includegraphics[width=8cm]{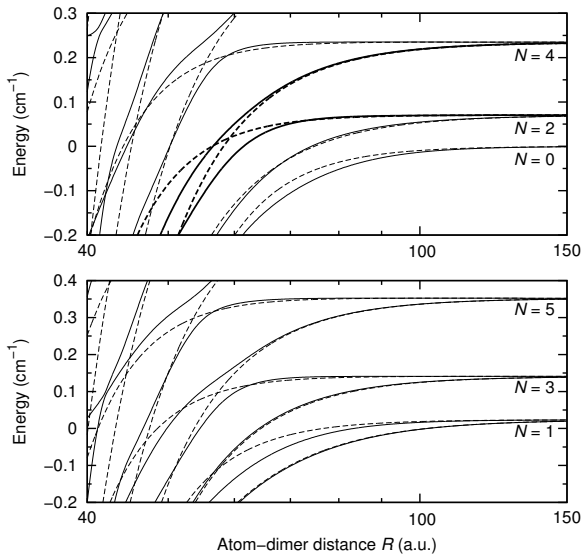}
\par\end{centering}
\caption{\label{fig:Ener-R-Sigma-p}Long-range potential energy curves between a ground-state Cs$_{2}$ and an excited Cs$(6^2P)$ atom, as functions of their mutual separation $R$, for the $\Sigma^{+}$ symmetry and for: (a) the even values of $N$ and (b) the odd values of $N$. The range of energy corresponds to the rotational energy of the dimer $B_{0}N(N+1)$ for $N=0$ to 5. The adiabatic curves, resulting from the diagonalization of $\hat{W}$ (see Eq. (\ref{eq:HamiltPertub})), are plotted in full lines. The diabatic ones, given by $B_{0}N(N+1)+C_{5}/R^{5}+C_{6}/R^{6}$ (see Paper II) are plotted in dashed lines. The curves in heavier lines are those involved in the crossing between $\left|(3)\Sigma^{+},N=2^{d}\right\rangle $ and $\left|(4)\Sigma^{+},N=4^{d}\right\rangle $, discussed in the text.}
\end{figure}
\begin{figure}
\begin{centering}
\includegraphics[width=8cm]{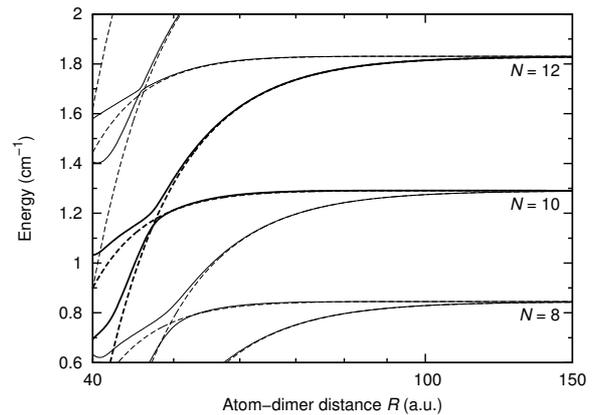}
\par\end{centering}
\caption{\label{fig:Ener-R-Sigma-p-Neleve}Same as Figure \ref{fig:Ener-R-Sigma-p}, except that the range of energy corresponds to the dissociation limits characterized by $N=8$ to 12. The curves in heavier lines are those involved in the crossing between $\left|(11)\Sigma^{+},N=10^{d}\right\rangle $ and $\left|(12)\Sigma^{+},N=12^{d}\right\rangle $, discussed in the text.}
\end{figure}
\begin{figure}
\begin{centering}
\includegraphics[width=8cm]{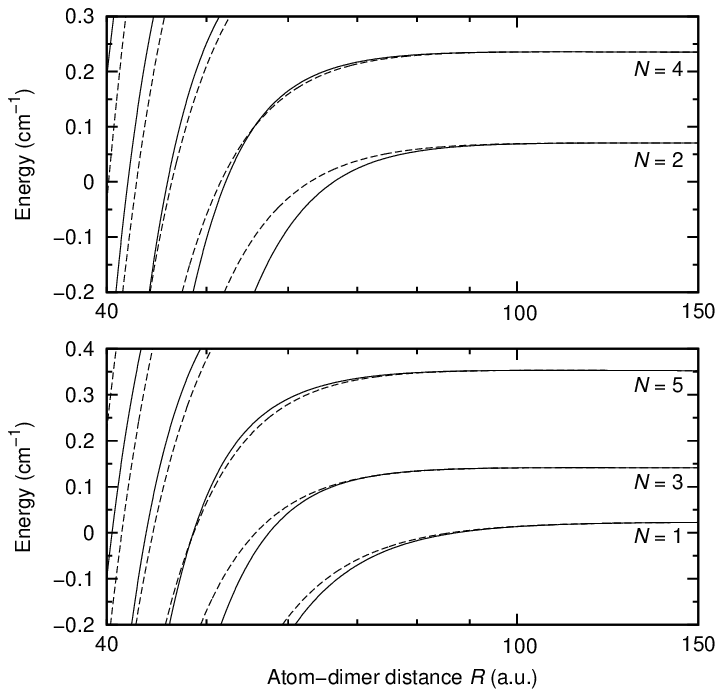}
\par\end{centering}
\caption{\label{fig:Ener-R-Sigma-m}Same as Figure \ref{fig:Ener-R-Sigma-p} but for the $\Sigma^{-}$ symmetry.}
\end{figure}
\begin{figure}
\begin{centering}
\includegraphics[width=8cm]{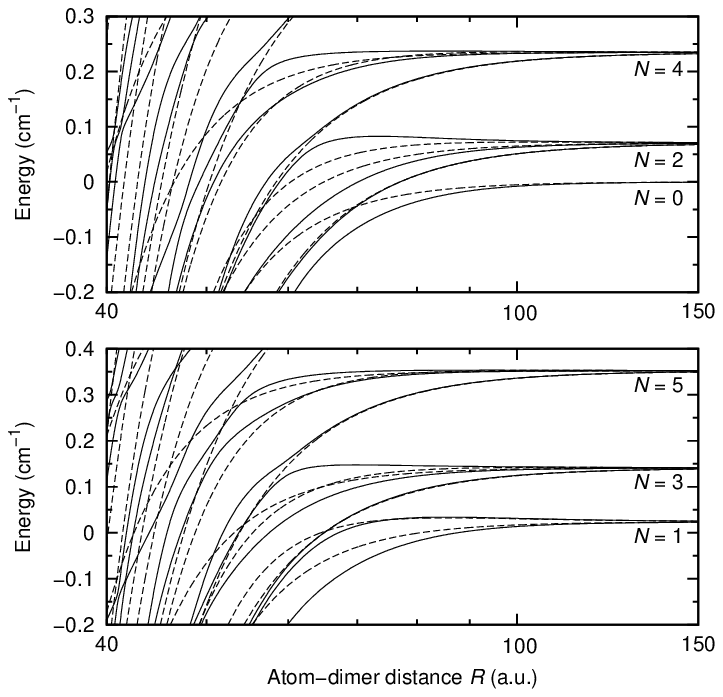}
\par\end{centering}
\caption{\label{fig:Ener-R-Pi}Same as Figure \ref{fig:Ener-R-Sigma-p} but for the $\Pi$ symmetry.}
\end{figure}
\begin{figure}
\begin{centering}
\includegraphics[width=8cm]{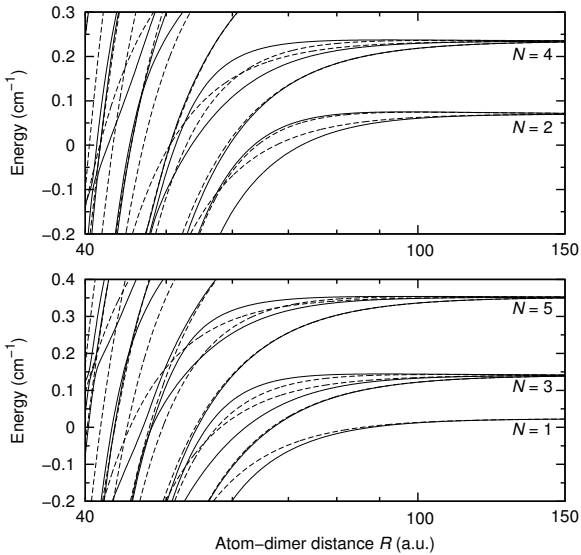}
\par\end{centering}
\caption{\label{fig:Ener-R-Delta}Same as Figure \ref{fig:Ener-R-Sigma-p} but for the $\Delta$ symmetry.}
\end{figure}
\begin{figure}
\begin{centering}
\includegraphics[width=8cm]{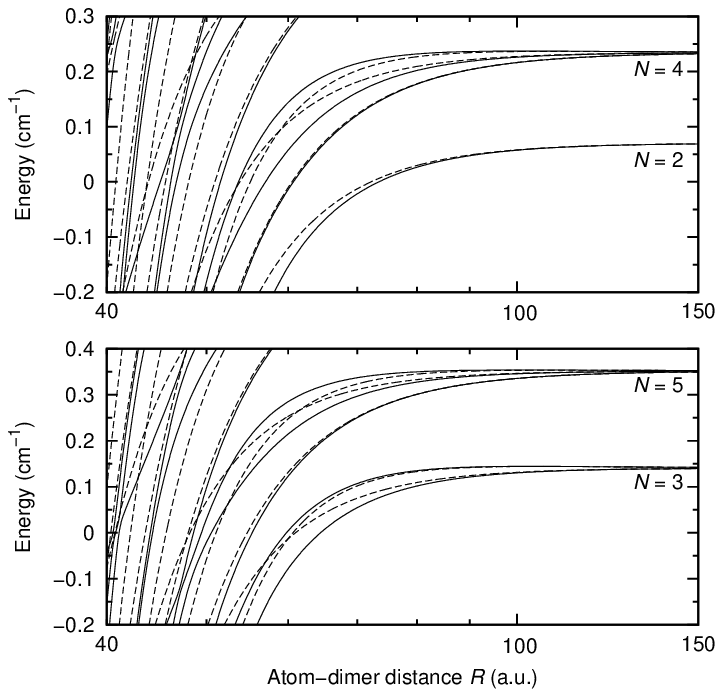}
\par\end{centering}
\caption{\label{fig:Ener-R-Phi}Same as Figure \ref{fig:Ener-R-Sigma-p} but for the $\Phi$ symmetry.}
\end{figure}
\begin{figure}
\begin{centering}
\includegraphics[width=8cm]{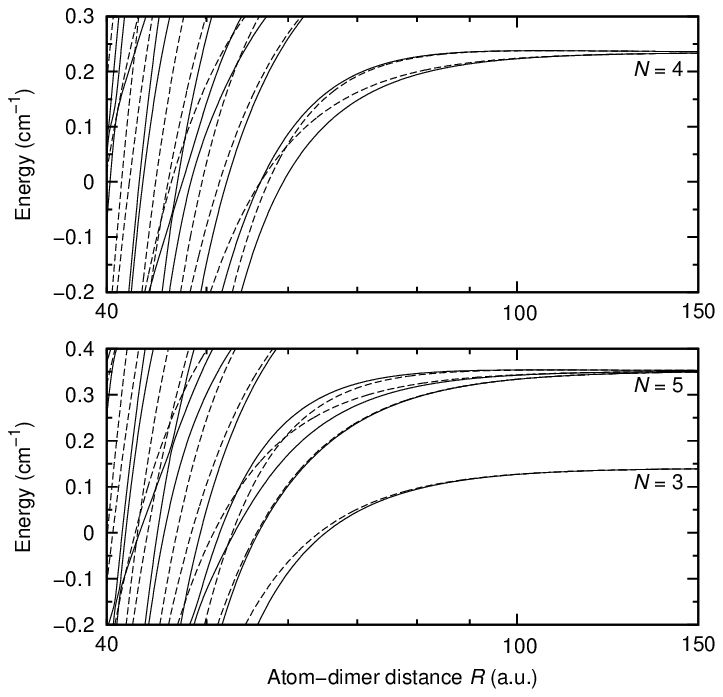}
\par\end{centering}
\caption{\label{fig:Ener-R-Gamma}Same as Figure \ref{fig:Ener-R-Sigma-p} but for the $\Gamma$ symmetry.}
\end{figure}
\begin{figure}
\begin{centering}
\includegraphics[width=8cm]{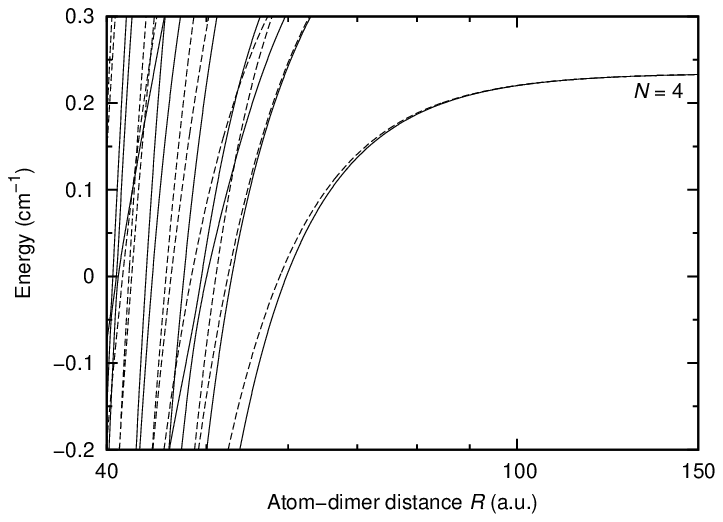}
\par\end{centering}
\caption{\label{fig:Ener-R-H}Same as Figure \ref{fig:Ener-R-Sigma-p} but for the H symmetry.}
\end{figure}

Both the diabatic and adiabatic potential energy curves are shown on Figs. \ref{fig:Ener-R-Sigma-p}--\ref{fig:Ener-R-H}, in dashed and solid lines respectively. Each panel corresponds to given $\left|m_{J}\right|$ and parity of $N$. Following our previous work, we have labelled our energy curves with the diatomic-like symmetries $\Sigma^{\pm}$, $\Pi$, $\Delta$, $\Phi$, $\Gamma$ and $H$. All the curves displayed on Figs. (\ref{fig:Ener-R-Sigma-p})--(\ref{fig:Ener-R-H}) present similar general features. On their right part, for $R\gtrsim100$ a.u., the diabatic and adiabatic curves are not distinguishable. This corresponds to the region of space in which the usual perturbation approach, characterized by the inequalities
\begin{equation}
\left\langle B_{Xv_{d}}\hat{\vec{N}}^{2}\right\rangle \gg\left\langle \hat{V}_{qq}\right\rangle \gg\left\langle \hat{V}_{dd}^{(2)}\right\rangle ,
\end{equation}
is applicable. In the central part, for $50\lesssim R\lesssim100$~a.u., strong differences are visible between the diabatic and adiabatic curves. In particular, we can see avoided crossings of the adiabatic curves, whereas the diabatic ones do cross. At last, in the left part, for $R\lesssim50$ a.u., we see a dense set of attractive curves connected to higher-$N$ dissociation limits.

Since the quadrupolar and dipolar parts of the Hamiltonian $\hat{W}$ (see Eqs. (\ref{eq:Vqq-MatElem}) and (\ref{eq:Vdd2-MatElem})) couple different values of $N$, each subspace of quasi-degeneracy has strictly speaking an infinite dimension. Of course, in our calculations, we have chosen a maximum value of $N$, $N_{max}$, and have tested the convergence with respect to that value. Our test indicate that, in order to describe properly the energy curves up to an energy of $B_{0}N^{*}(N^{*}+1)$, one has to chose at least $N_{max}\approx N^{*}+4$. For our calculations, we have taken $N_{max}=17$, which gives us a reliable description of the range of energy in Fig. \ref{fig:Ener-R-Sigma-p-Neleve}. This observation can be explained as follows. In the range of distances in which we are interested, there exists a value of $R$ where the most attractive curve coming from the $N^{*}+2$ dissociation limit has the energy $B_{0}N^{*}(N^{*}+1)$ corresponding to the $N^{*}$ dissociation limit. So, in order to correctly describe that $N^{*}+2$ curve, one has to include in the calculations the curves connected to the $N^{*}+4$ dissociation limit, as they are mutually coupled. By ignoring the $C_{6}/R^{6}$ part of the energy of the $N^{*}+2$ curve, we can estimate that the crossing region scales as $R\sim\left(4N+6\right)^{-1/5}$. This weak sensitivity to $N$ explains why the difference $N_{max}-N^{*}$ does not vary significantly with increasing $N^{*}$, although the rotational levels of the dimer get further away from each other. This is particularly the case for the low energies considered here, which are characterized by $N^{*}\le20$.

It is worthwhile noting that the previous reasoning does not hold for $R\le40$~a.u.. In this region, the convergence on $N_{max}$ is very difficult to obtain. Due to the very strong $R$-dependence of the electrostatic interaction, the influence of the $N^{*}\pm2$ curves is not the only one significant on the $N^{*}$ curves. We have not represented such a region on Figs. \ref{fig:Ener-R-Sigma-p}--\ref{fig:Ener-R-H}, since the electronic overlap will certainly play an important role. However, it cannot be excluded that such a convergence problem arises with different reactants.

In the region $40\lesssim R\lesssim100$~a.u., the interactions between different diabatic curves strongly modify the aspect of the potential energy curves. The most obvious modifications are the clear avoided crossings of the adiabatic curves. This is for example the case between the $\left|(3)\Sigma^{+},N=2^{d}\right\rangle $ and the $\left|(4)\Sigma^{+},N=4^{d}\right\rangle $ states, or between the $\left|(11)\Sigma^{+},N=10^{d}\right\rangle $ and the $\left|(12)\Sigma^{+},N=12^{d}\right\rangle $ (see Figs. (\ref{fig:Ener-R-Sigma-p}) and (\ref{fig:Ener-R-Sigma-p-Neleve}) respectively). More generally, for the $\Sigma^{+}$ symmetry, there is an avoided crossing between the less attractive curve of the $N$ manifold and the more attractive one in the $N+2$ manifold. In non-$\Sigma$ symmetries, the higher density of states (3 for each $N$ instead of 2 for $\Sigma^{+}$) alters the usual two-state picture of the Landau-Zener problem, due to the influence of neighboring curves. At last, for the $\Sigma^{-}$ symmetry, we cannot see any Landau-Zener crossing, as the diabatic curves are less numerous and approximately parallel.

We have estimated the Landau-Zener probability $P_{pr}$ to go through the crossing between the curves $(p)$ and $(r)$ in a diabatic way, and applied the result to the two crossings evoked above. We start from the general formula
\begin{equation}
P_{pr}=\exp\left(-2\pi\Gamma_{pr}\right),
\label{eq:p-lz}
\end{equation}
with
\begin{equation}
\Gamma_{pr}=\frac{\left|W_{pr}^{d}(R_{0})\right|^{2}}{v_{p}(R_{0},T)\left|\frac{\partial}{\partial R}W_{pp}^{d}(R=R_{0})-\frac{\partial}{\partial R}W_{rr}^{d}(R=R_{0})\right|}\,.
\end{equation}
In Eq. (\ref{eq:p-lz}), $W_{pr}^{d}(R)$ is the matrix element of the hamiltonian (\ref{eq:HamiltPertub}) in the diabatic representation
\begin{equation}
W_{pr}^{d}(R)=\left\langle (p)m_{J1}N_{1}^{d}\left|\hat{W}(R)\right|(r)m_{J2}N_{2}^{d}\right\rangle ,
\label{eq:Diab-MatrElem-W}
\end{equation}
$R_{0}$ is the distance at which the crossing occurs, and $v(R,T)$
is the temperature-dependent classical velocity of a particle entering the crossing in the $p$ channel,
\begin{equation}
v_{p}(R,T)=\sqrt{\frac{2\left(W_{pp}(R\to\infty)-W_{pp}(R)\right)+3k_{B}T}{\mu}}\,,
\end{equation}
with $k_{B}$ Boltzmann's constant and $\mu$ the atom-dimer reduced
mass. The off-diagonal
term $W_{pr}(R)$ appearing in Eq. (\ref{eq:p-lz}) can be written
\begin{equation}
W_{pr}(R)=\frac{C_{5}'}{R^{5}}+\frac{C_{6}'}{R^{6}}
\end{equation}
where $C_{5}'$ and $C_{6}'$ are the off-diagonal matrix elements:
$C_{5}'=R^{5}\left\langle p\left|\hat{V}_{qq}\right|r\right\rangle $
and $C_{6}'=R^{6}\left\langle p\left|\hat{V}_{dd}^{(2)}\right|r\right\rangle $;
and the derivative $\frac{\partial}{\partial R}W_{pp}(R)$ reads
\begin{equation}
\frac{\partial}{\partial R}W_{pp}=-\frac{5C_{5}}{R^{6}}-\frac{6C_{6}}{R^{7}}.
\end{equation}

\begin{table}
\begin{centering}
\begin{tabular}{|c|c|c|}
\hline
entrance channel $p$ & $v_p$ ($10^{-6}\times$a.u.) & $P_{pr}\,(\%)$ \tabularnewline
\hline
$\left|(3)\Sigma^{+},N=2^d\right\rangle$ & 2.4 & 4 \tabularnewline
$\left|(4)\Sigma^{+},N=4^d\right\rangle$ & 4.9 & 21 \tabularnewline
\hline
\end{tabular}
\par\end{centering}
\caption{\label{tab:p-lz-1}Characterization of the Landau-Zener crossing between the $\left|(3)\Sigma^{+},N=2^d\right\rangle$ and $\left|(4)\Sigma^{+},N=4^d\right\rangle$ diabatic curves, with respect to each entrance channel. The crossing occurs at $R_0=59.3$~a.u., and the corresponding coupling $W_{pr}$ is 0.058~cm$^{-1}$. The velocities $v_p$ are calculated for a temperature $T=1$~mK.}
\end{table}

\begin{table}
\begin{centering}
\begin{tabular}{|c|c|c|}
\hline
entrance channel $p$ & $v_p$ ($10^{-6}\times$a.u.) & $P_{pr}\,(\%)$ \tabularnewline
\hline
$\left|(11)\Sigma^{+},N=10^d\right\rangle$ & 3.6 & 89 \tabularnewline
$\left|(12)\Sigma^{+},N=12^d\right\rangle$ & 8.6 & 95 \tabularnewline
\hline
\end{tabular}
\par\end{centering}
\caption{\label{tab:p-lz-2}Characterization of the Landau-Zener crossing between the $\left|(11)\Sigma^{+},N=10^d\right\rangle$ and $\left|(12)\Sigma^{+},N=12^d\right\rangle$ diabatic curves, with respect to each entrance channel. The crossing occurs at $R_0=47.6$~a.u., and the corresponding coupling $W_{pr}$ is -0.027~cm$^{-1}$. The velocities $v_p$ are calculated for a temperature $T=1$~mK.}
\end{table}

We have applied Eq. (\ref{eq:p-lz}) to the crossings between $\left|(3)\Sigma^{+},N=2^d\right\rangle$ and $\left|(4)\Sigma^{+},N=4^d\right\rangle$, and between $\left|(11)\Sigma^{+},N=10^d\right\rangle$ and $\left|(12)\Sigma^{+},N=12^d\right\rangle$, and we have summarized the results in Tables \ref{tab:p-lz-1} and \ref{tab:p-lz-2}, respectively. The dynamics can drastically change from one case to another, turning from almost adiabatic in the lower channel of the former crossing ($P_{pr}=4\,\%$), to almost diabatic in the latter crossing ($P_{pr}=95\,\%$). Obviously, for a given crossing, the dynamics of a particle coming from the upper channel is more diabatic, as it has acquired more kinetic energy at the crossing point. However, in the crossing between the (11) and (12) curves, even if the velocity of the upper channel is more than twice the velocity of the lower one, the Landau-Zener probability only increases by $6 \%$. Actually, the most determinant parameter for the nature of the crossing is the coupling matrix element $W_{pr}^{d}$. The second crossing tends to be more diabatic, because $\left|W_{pr}^{d}\right|$ is significantly smaller, \textit{i.e.} -0.027 cm$^{-1}$ with respect to 0.058 cm$^{-1}$. This shrinking is due to the competition between $C'_{5}>0$ and $C'_{6}<0$.

Apart from the Landau-Zener-like crossings, the diabatic curves can also change their behavior in a {}``smoother'' way as a function of $R$. This is for example the case for the $(1)\Sigma^{+}$ and $(1)\Pi$ states correlated to $N=0$, and for the $(2)\Sigma^{+}$ and $(2)\Pi$ states correlated to $N=1$. For $R\ge100$~a.u., those states are characterized by $C_{5}=0$; but as $R$ decreases, they progressively acquire a $R^{-5}$ behavior, due to the coupling with the higher curves. This result might be of strong importance, especially in the prospect of creating an excited trimer from a $N=0$ sample of dimers.

This variety of coupling situations has made us searching for a better criterion of applicability of the $C_{n}/R^{n}$ expansion, than the one proposed in Paper I, and based of the curve crossings. Basically, we estimate that the $C_{n}/R^{n}$ expansion stops to be valid as soon as the ratio between the coupling and the energy difference between two states exceeds a certain threshold $\epsilon$. Therefore we define a state-to-state normalized coupling $\overline{W}_{pr}$ as
\begin{equation}
\overline{W}_{pr}(R)=\frac{W_{pr}^{d}(R)}{W_{pp}^{d}(R)-W_{rr}^{d}(R)},
\label{eq:cpling-wbar}
\end{equation}
where $W_{pr}^{d}(R)$ are the matrix elements of $\hat{W}$ in the diabatic representation, and we consider that the diabatic representation fails for $R$ such that
\begin{equation}
\max_{r}\left|\overline{W}_{pr}(R)\right|\ge\epsilon.
\label{eq:criere-adiab}
\end{equation}
In Eq. (\ref{eq:criere-adiab}), the equality corresponds to the threshold distance $R=R_{p}^{*}$, which depends on the state $\left|(p)m_{J}N^{d}\right\rangle $ under consideration.
\begin{figure}
\begin{centering}
\includegraphics[width=8cm]{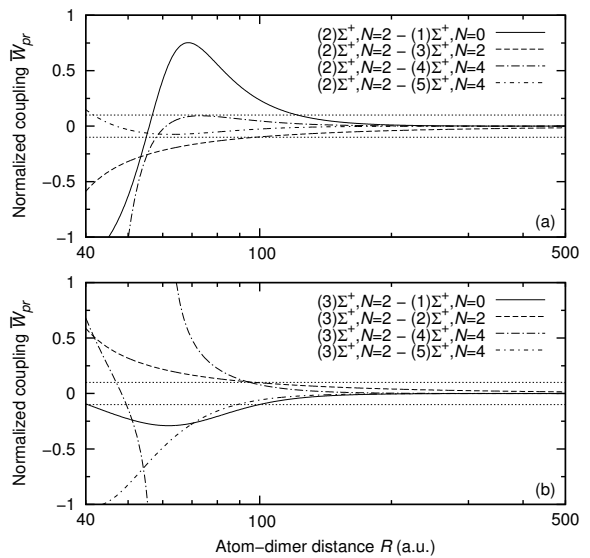}
\par\end{centering}
\caption{\label{fig:Cpling-R-Sigma-p}Normalized state-to-state couplings $\overline{W}_{pr}$ (see Eq. (\ref{eq:cpling-wbar})) as functions of $R$, for the two diabatic states of the $\Sigma^{+},N=2$ manifold: (a) state (2), (b) state (3). The two horizontal dotted lines correspond to $\left|\overline{W}_{pr}\right|=\epsilon=0.1$.}
\end{figure}
\begin{figure}
\begin{centering}
\includegraphics[width=8cm]{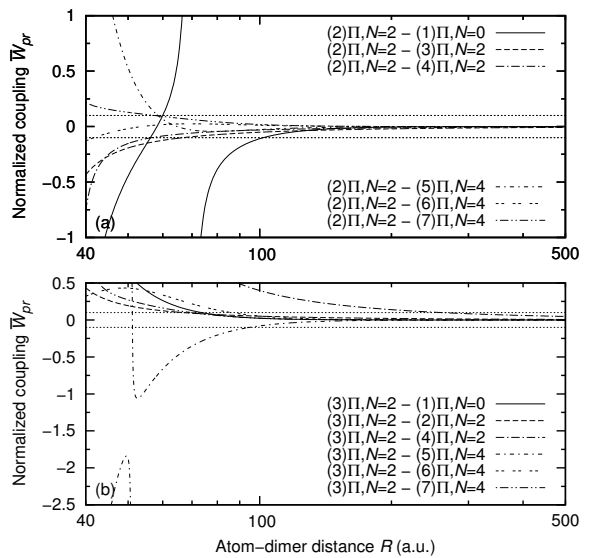}
\par\end{centering}
\caption{\label{fig:Cpling-R-Pi}Same as Fig. \ref{fig:Cpling-R-Sigma-p}, but for two diabatic states of the $\Pi,N=2$ manifold: (a) state
(2), (b) state (3).}
\end{figure}
\begin{figure}
\begin{centering}
\includegraphics[width=8cm]{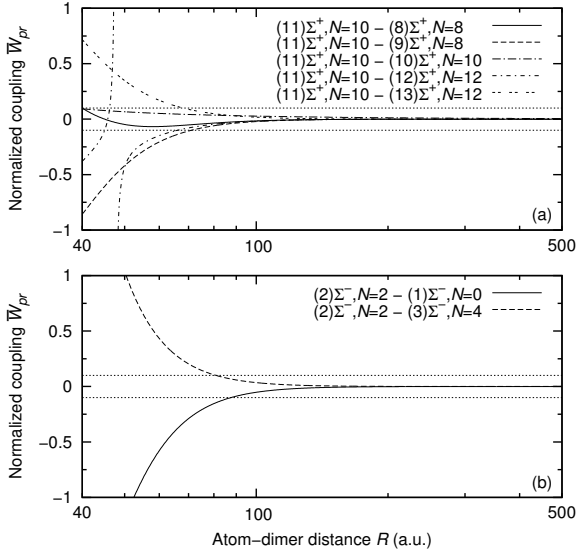}
\par\end{centering}
\caption{\label{fig:Cpling-R-Sigma}Same as Fig. \ref{fig:Cpling-R-Sigma-p} but: (a) for state $\left|(11)\Sigma^{+},N=10\right\rangle $, and (b) state $\left|(2)\Sigma^{-},N=2\right\rangle $.}
\end{figure}
The quantity $\overline{W}_{pr}$ is plotted on Figs. \ref{fig:Cpling-R-Sigma-p}-\ref{fig:Cpling-R-Sigma} for various symmetries. In all cases, the quantities $\overline{W}_{pr}$ tend to zero with increasing atom-dimer distances (note that $R$ goes up to 500 a.u. on Fig. \ref{fig:Cpling-R-Sigma-p}-\ref{fig:Cpling-R-Sigma}), since the off-diagonal elements of $\hat{W}$ vanish. On the contrary, $\overline{W}_{pr}$ increases with smaller $R$. For given pairs of states, $\overline{W}_{pr}$ even shows a resonant behavior, corresponding to a crossing of the diabatic curves. For example, we observe the crossings already mentioned above, between $\left|(3)\Sigma^{+},N=2\right\rangle $ and $\left|(4)\Sigma^{+},N=4\right\rangle $, and between $\left|(11)\Sigma^{+},N=10\right\rangle $ and $\left|(12)\Sigma^{+},N=12\right\rangle $. We note the sharpness of the latter resonance compared to the former one, which confirms its more diabatic character. We also note two resonances between $\left|(1)\Pi,N=0\right\rangle $ and $\left|(2)\Pi,N=2\right\rangle $, reflecting two intersections of those two curves at $R=32$ and 69 a.u.. Another surprising aspect is that of the $\left|(3)\Pi,N=2\right\rangle $-$\left|(5)\Pi,N=4\right\rangle $ resonance, due to the fact that the numerator and the denominator of $\overline{W}_{pq}(R)$ both vanish at $R\approx50$~a.u..

On each graph of Figs. \ref{fig:Cpling-R-Sigma-p}-\ref{fig:Cpling-R-Sigma}, we also display two horizontal lines characterized by $\left|\overline{W}_{pr}\right|=10\%$, that we consider as the limit of applicability of the $C_{n}/R^{n}$ expansion. This criterion gives different low-$R$ limits $R_{p}^{*}$ according the state under consideration. It is quite similar for the states $\left|(2)\Sigma^{+},N=2\right\rangle $, $\left|(3)\Sigma^{+},N=2\right\rangle $ and $\left|(2)\Pi,N=2\right\rangle $, respectively equal to 122, 100 and 102~a.u., and also quite close to the value $R_{m}=102$~a.u. determined in another way in Paper I. For state $\left|(3)\Pi,N=2\right\rangle $, all the crossings take place in the same region, except one, with state $\left|(4)\Pi,N=2\right\rangle $, for which $R_{p}^{*}=261$~a.u.. The very slow convergence of the corresponding $\overline{W}_{pr}(R)$ function can be explained as follows. The two states, that belong to the $N=2$ manifold are characterized by a small energy difference: \emph{i.e.} at $R=200$~a.u., $3.3\times10^{-4}$ cm$^{-1}$, compared to $-5.4\times10^{-2}$ cm$^{-1}$ between (3) and (2), due to a compensation effect between the $C_{5}$ and $C_{6}$ coefficients. By contrast, they are still coupled at very long distance due to the dipolar interaction, equal to $4.4\times10^{-4}$ cm$^{-1}$ compared to $-1.1\times10^{-4}$ cm$^{-1}$ between (3) and (2) at 200 a.u.. The particular occurrence can also be observed for higher $N$, \emph{e.g.} the couple $\left|(6)\Pi,N=4\right\rangle $-$\left|(7)\Pi,N=4\right\rangle $, but it tends to damp as $N$ increases. Similar features have been noted in the case of two interacting atoms \cite{rerat1997}.

At last, Fig. \ref{fig:Cpling-R-Sigma} illustrates two situations where the value $R_{p}^{*}$ is significantly smaller. On panel (a), we have plotted $\overline{W}_{pr}(R)$ for the state $\left|(11)\Sigma^{+},N=10\right\rangle $. As shown on Fig. \ref{fig:Ener-R-Sigma-p-Neleve}, the diabatic analysis is estimated to be valid down to $R_{p}^{*}=71$ a.u.. The conclusion is similar for the $\left|(2)\Sigma^{-},N=2\right\rangle $ represented on panel (b), since $R_{p}^{*}=88$ a.u..

\section{Conclusion and prospects}
\label{sec:conclusion}

In this article, we have characterized the long-range interactions
between an alkali-metal dimer in the electronic ground state and arbitrary
vibrational and rotational levels, and an excited alkali-metal atom.
More precisely, we have studied in details the interactions due to
the permanent quadrupoles and to the induced dipoles of the two reactants.
We have focused ourselves in a range of atom-dimer distances where
the electrostatic energy competes with the rotational structure of
the dimer, which allows couplings between the different rotational
levels.

The potential energy curves that we have obtained are characteristic
of coupled-channel problems, showing in particular avoided crossings.
By calculating the Landau-Zener probability for some of those crossings,
we have seen that the dynamics can strongly turn from adiabatic to
diabatic, when the dimer rotational quantum number increases. Such
characteristics have already been predicted with the $[\textrm{OH}]_{2}$
complex \cite{avdeenkov2002,avdeenkov2003} and with polar $^{1}\Sigma$
molecules \cite{avdeenkov2006}, submitted to an external electric field.

Because of the strongly attractive induced-dipole interaction, all
the curves displayed in this article are attractive, which is an encouraging
result in the prospect of forming excited trimers by photoassociation.
However, we note the absence of long-range potential wells, that can
have a key role in the first observation of cold Cs$_2$ molecules \cite{fioretti1998}

We have illustrated our formalism in the particular case of a Cs$_{2}$
dimer and a cesium atom. But we expect our conclusions to be generalizable
to other atom-dimer systems and even to dimer-dimer systems, except
hydrides, since the balance between the rotational and electrostatic
energies does not strongly vary for other alkali-metals. For example,
in the case of Li$_{2}+$Li$(2^{2}P)$, the dimer rotational constant
is about 60 times higher than that of Cs$_{2}$, and the quadrupole
moments and polarizabilities are smaller. But, since the electronic
overlap becomes important at smaller distances, we can predict the existence
of a crossing region, that we have estimated between 26 and 43 u.a.
\cite{lepers2010}.

By contrast, the inclusion of the atomic fine structure will create
a species-dependent situation. In the case of lithium, the spin-orbit
splitting of the $2^{2}P$ state being 0.335 cm$^{-1}$, it is comparable
to the typical rotational and electrostatic energies, and may therefore
complicate the potential curves displayed in this article. On the
contrary, for the other alkali-metal atoms, the spin-orbit splitting
is so high (e.g. 17 cm$^{-1}$ for Na$(3^{2}P)$), that the two fine-structure
components $nP_{1/2}$ and $nP_{3/2}$ will not be coupled by the
electrostatic interaction.

\section*{Acknowledgements}
Stimulating discussions with N. Bouloufa, V. Kokoouline, and R. Vexiau, are gratefully acknowledged. M. L. acknowledges the support of \textit{Triangle de la Physique} in the framework of the contract QCCM-2008-007T \textit{Quantum Control of Cold Molecules}.
%\bibliography{../DATABASE/bibliocold}
%\bibliographystyle{epj}

\end{document}